\documentclass{article}

\PassOptionsToPackage{numbers, compress}{natbib}

\usepackage{ifthen}
\usepackage[dvipsnames]{xcolor}

\newboolean{ANONYMIZE}
\setboolean{ANONYMIZE}{false}  

\ifthenelse{\boolean{ANONYMIZE}}{
  \newcommand{\mawad}[1]{}
  \newcommand{\ryan}[1]{}
  \newcommand{\ganesh}[1]{}
}{
  \newcommand{\mawad}[1]{\textcolor{blue}{\textbf{\textit{mawad: #1}}}}
  \newcommand{\ryan}[1]{\textcolor{teal}{\textbf{\textit{ryan: #1}}}}
  \newcommand{\ganesh}[1]{\textcolor{red}{\textbf{\textit{ganesh: #1}}}}
}

\newcommand{\nonbijective}{{\large\textcircled{\footnotesize$\times$}}}

\newcommand{\rocprof}{\texttt{rocprofv3}} 

\usepackage[preprint]{swizzleperf}

\usepackage[utf8]{inputenc} 
\usepackage[T1]{fontenc}    
\usepackage[hidelinks]{hyperref}       
\usepackage{url}            
\usepackage{booktabs}       
\usepackage{amsfonts}       
\usepackage{nicefrac}       
\usepackage{microtype}      
\usepackage[most]{tcolorbox}   
\usepackage{ragged2e}          
\usepackage{setspace}          
\usepackage{graphicx}     
\usepackage{subcaption}   

\usepackage{caption}
\usepackage{float}
\usepackage{listings}
\usepackage{caption}

\title{SwizzlePerf: Hardware-Aware LLMs for\\GPU Kernel Performance Optimization}

\author{%
  \textbf{Arya Tschand}$^{1}$\thanks{Work done as an intern at AMD.} \quad
  \textbf{Muhammad Awad}$^{2}$ \quad
  \textbf{Ryan Swann}$^{2}$ \quad
  \textbf{Kesavan Ramakrishnan}$^{3}$ \\
  \textbf{Jeffrey Ma}$^{1}$ \quad
  \textbf{Keith Lowery}$^{2}$ \quad
  \textbf{Ganesh Dasika}$^{2}$ \quad
  \textbf{Vijay Janapa Reddi}$^{1}$ \\[6pt]
  $^{1}$Harvard University \quad
  $^{2}$AMD \quad
  $^{3}$Stanford University \\[6pt]
  \texttt{aryatschand@g.harvard.edu}$^{1}$ \\
}

\begin{document}

\maketitle

\vspace{-16pt}
\begin{abstract}
  Large language models (LLMs) have shown progress in GPU kernel performance engineering using inefficient search-based methods that optimize around runtime. Any existing approach lacks a key characteristic that human performance engineers rely on for near-optimal utilization -- \textit{hardware-awareness}. By leveraging the workload's specific memory access patterns, architecture specifications, filtered profiling logs, and reflections on historical performance, we can make software-level optimizations that are tailored to the underlying hardware. SwizzlePerf automatically generates spatial optimizations for GPU kernels on disaggregated architectures by giving LLMs explicit hardware-awareness. 
  \vspace{8pt}

For a GEMM kernel, SwizzlePerf takes less than 5 minutes to generate the same hardware-specific optimal swizzling pattern that took expert performance engineers 2 weeks to find. On a suite of 10 diverse ML and Science kernels, SwizzlePerf can generate swizzling patterns for 9 of the kernels that achieve up to a $2.06 \times$ speedup and $70\%$ improvement in L2 hit rate. This work is the first of many steps toward systematically creating hardware-aware LLM performance engineering agents.


\end{abstract}
\vspace{-12pt}
\section{Introduction}
\vspace{-2pt}
    GPU code performance engineering is a necessary step in enabling efficient machine learning (ML) systems and High-Performance Computing (HPC) applications. Performance engineering requires hardware-software codesign through understanding how the specific workload is executing on the specific underlying hardware. In this work, we automate this process for GPU kernels by imitating this hardware-software codesign process that human performance engineers follow. This is accomplished by giving LLMs hardware-aware context, which unlocks their ability to structure GPU code optimizations around the underlying hardware architecture and scheduling methodology.

\begin{figure}[H]
  \centering

  \subcaptionbox{Optimization loop methodology.\label{fig:swizzleperf}}[0.72\linewidth]{%
    \includegraphics[width=\linewidth]{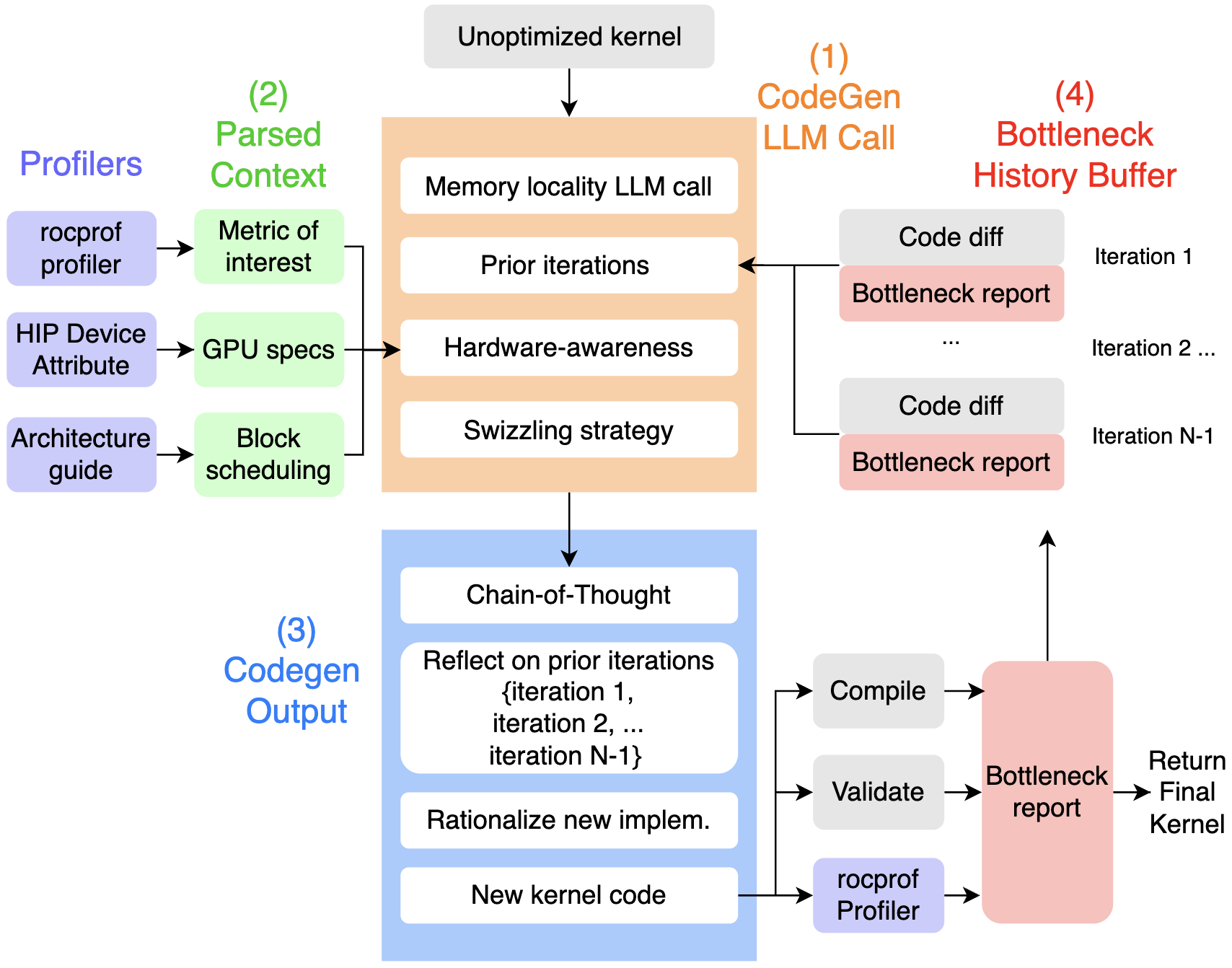}
  }
  \hfill
  \subcaptionbox{GEMM swizzling.\label{fig:gemmswizzle}}[0.25\linewidth]{%
    \includegraphics[width=0.85\linewidth]{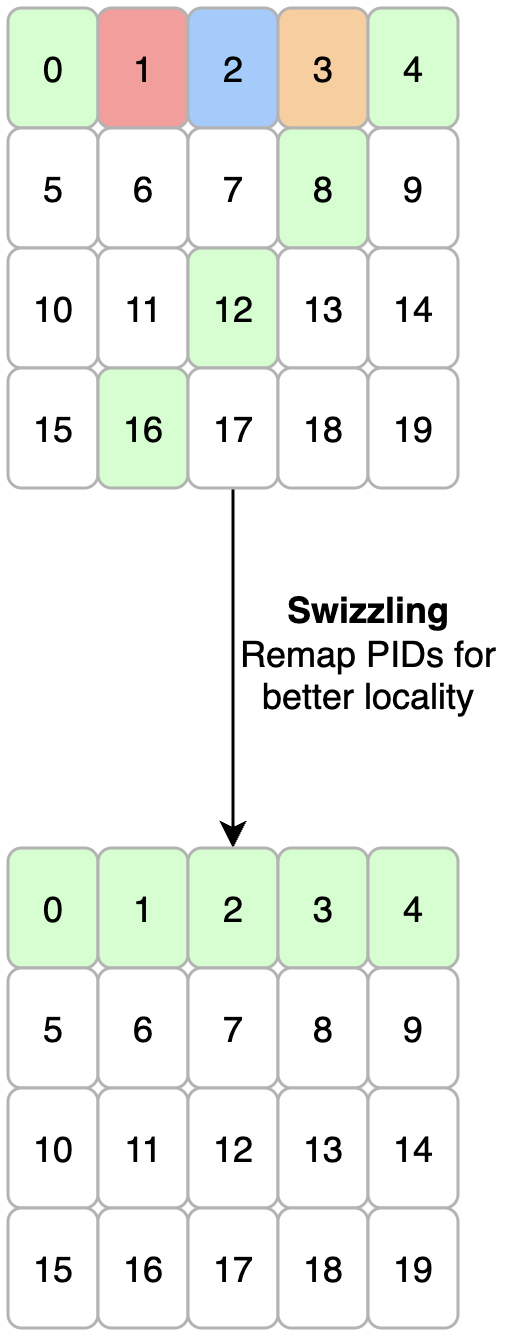}
  }
\caption{\emph{SwizzlePerf methodology and example swizzling outcome.} 
\textbf{(a)} The optimization loop begins with the \textcolor{orange}{CodeGen LLM call input}, which integrates the \textcolor{ForestGreen}{parsed context} of bottleneck metrics, GPU characteristics, and the scheduling policy. 
The LLM reflects on \textcolor{red}{past bottlenecks} and proposes a new swizzling formula in the \textcolor{blue}{CodeGen LLM call output}.
\textbf{(b)} SwizzlePerf generates this pattern for GEMM on a 4-XCD architecture. The swizzling pattern improves locality in the L2 cache by co-locating tiles that reuse rows in A on the same XCD.
While the goal is always to improve intra-XCD locality, the exact swizzling formula varies by algorithm.}
\end{figure}

    \textit{SwizzlePerf} is our proposed hardware-aware LLM workflow that automatically generates \textit{swizzling} patterns. Swizzling is a transformation that reorders the mapping between data or work and their execution/storage locations to enhance spatial/temporal locality and align with hardware topology. Compilers and runtimes use swizzling for data layouts and for clustering cooperating blocks. In the case study that we evaluate in this paper, swizzling is a programmer-defined remapping of GPU workgroup program IDs (PIDs) that co-locates related tiles on the same accelerator complex die (XCD) to increase per-XCD L2 reuse on GPUs with disaggregated architectures. By default, workgroups are assigned in a round-robin manner across XCDs (Figure~\ref{fig:gemmswizzle} and Appendix~\ref{appendix_gemm}). For applications with predictable memory-access patterns, statically arranging computation and data to maximize local L2 access yields substantial benefits. 

    It takes expert GPU performance engineers multiple weeks to make targeted spatial optimizations. SwizzlePerf autonomously finds these optimizations for a wide range of kernels in minutes. 
    
\vspace{4pt}
\textbf{Contributions of this work:}
    \begin{enumerate}



    
        \item We facilitate hardware-aware kernel optimization by intentionally crafting context with necessary hardware and scheduling information from profilers and runtimes.
        \item We augment IntelliPerf, an open-source autonomous performance engineering tool, with hardware-awareness and evaluate a case study on generating GPU kernel swizzling patterns.
        \item We show performance results on a wide range of ML and scientific GPU kernels of up to a 2.1x speedup and 70\% L2 hit rate improvement, indicating that hardware-awareness is necessary to unlock hardware-specific optimizations.
        \end{enumerate}

\vspace{-4pt}
\vspace{-4pt}
\vspace{-4pt}
\vspace{-4pt}
\vspace{-4pt}
\vspace{-4pt}
\section{Related Work}
\vspace{-4pt}

Prior work on autonomous kernel optimization largely optimizes for \textit{runtime} as the single objective using three families of methods: (i) heuristic autotuners and analytic cost models that apply rule-based schedules (tiling, unrolling, coalescing)~\cite{whaley2001atlas,frigo2005fftw3,ansel2014opentuner} (ii) learned cost models that predict runtime from code or schedule features and drive combinatorial search~\cite{chen2018learning,zheng2020ansor,adams2019halide} and (iii) LLM/RL-driven search that performs test-time exploration or multi-turn refinement, ranging from competitive-programming systems~\cite{alphacode, pie4perf} to GPU-kernel frameworks~\cite{kernelbench, kevin, andrews2025gpu, li2025cuda}. These approaches can work for software-level bottlenecks, but they typically abstract away concrete architectural constraints (cache topology, chiplet boundaries, block-placement policy) and provide minimal task-specific machine feedback. Without hardware-awareness, the search space becomes too broad and the objective (runtime) is noisy due to launch overheads, overlapping bottlenecks, and compute saturation. To converge on \textit{hardware-specific} optimizations, the LLM needs relevant context that enables \textit{hardware-awareness}.

We approach this problem by imitating human performance engineers: profile the workload, isolate the bottleneck, apply a targeted fix, and precisely evaluate the improvement. Practically, this means intentionally supplying the profiling, architectural, and scheduling context that is relevant to the optimization task and guiding feedback with a \textit{bottleneck metric} that provides a stable signal aligned with the transformation. By elevating this metric to the core optimization objective, we narrow the search to changes that directly target the bottleneck. For the cache efficiency bottleneck case study in this paper, we measure L2 hit rate as a direct, low-noise proxy metric for spatial locality. To our knowledge, SwizzlePerf is the only work that adds rich context from a suite of profilers into the context to directly reflect cache-locality improvements and improve LLM optimization.

Within this framing, we implement \textit{swizzling} patterns for disaggregated GPUs (e.g., AMD Instinct\textsuperscript{TM} MI300x \cite{mi300x}) where multiple accelerator complex dies (XCDs) each host a dedicated L2 cache. By default, workgroups are scheduled round-robin across XCDs (Figure~\ref{fig:gemmswizzle}, Appendix~\ref{appendix_gemm}). However, for applications with predictable access patterns, statically co-locating cooperating tiles on the same XCD with swizzling can substantially improve data reuse in the shared L2 cache. Our contribution adds hardware-awareness and optimizes around a bottleneck metric so the search is focused, the signal is stable, and the resulting transformations align with the underlying hardware.

\lstset{%
  basicstyle=\ttfamily\footnotesize,
  breaklines=true,
  columns=fullflexible,
  frame=single,
  aboveskip=0pt,   
  belowskip=0pt    
}

\vspace{-6pt}

\section{Methodology}

\vspace{-2pt}

SwizzlePerf (Figure~\ref{fig:swizzleperf}) is a hardware-aware, bottleneck-driven optimization loop.

\textcolor{orange}{\textbf{(1) CodeGen LLM Call.}} We formulate a targeted code-generation request that fixes the optimization objective and explicitly defines the bottleneck metric of interest. The prompt (Appendix \ref{appendix_prompts}) bundles a short memory-locality summary of the kernel, a compact trace of prior attempts, and architecture details (e.g., number of XCDs, cache sizes, and the block-scheduling policy). We build on the open source IntelliPerf ~\cite{Awad:2025:ILP} framework that runs a \textit{hardware-unaware} GPU kernel optimization loop for a single bottleneck. We constrain the goal to a specific, hardware-aware transformation.

\textcolor{ForestGreen}{\textbf{(2) Parsed Context.}} We construct a structured context from public profilers and documentation. From \rocprof{}\cite{AMD:2025:RAP} we extract bottleneck metrics, from HIP device attributes\cite{AMD:2025:HIP} we gather GPU/XCD/cache parameters, and from architecture guides we derive the default block-scheduling policy. This parsed context exposes the metric of interest (for swizzling, this metric is L2 hit rate) and the spatial constraints the remapping must satisfy. Related work does not include any profiler-provided information in context, and IntelliPerf only profiles the output code for the bottleneck metric.

\textcolor{blue}{\textbf{(3) CodeGen Output.}} Using DSPy~\cite{Khattab:2022:DSP, Khattab:2024:DCL}, we specify the output signature so the model must critique past attempts and propose a new swizzling pattern. The LLM returns (i) a reasoning trace that contrasts old and new mappings, and (ii) a swizzling formula implementation. We then compile the new code, validate correctness against the ground truth, and run \rocprof{} to obtain the updated bottleneck report. We reuse IntelliPerf’s compile/validate/profile scaffold, but the code generation is guided by hardware-aware context and a fixed objective rather than unconstrained search.

\textcolor{red}{\textbf{(4) Bottleneck History Buffer.}} Each iteration appends the code diff and bottleneck report to a persistent history buffer. Subsequent calls see this history, reflect on failures (e.g., broken mappings or no L2 change), and propose diversified remappings. We rank candidates by L2 hit rate (primary) and retain the best validated kernel. This buffer closes the hardware–software loop by feeding back a bottleneck-specific signal that accelerates convergence to architecture-aligned swizzling patterns.

\vspace{6pt}

 \vspace{-16pt}
\section{Results}
 \vspace{-4pt}

\begin{figure}[H]
\centering
\includegraphics[width=\linewidth]{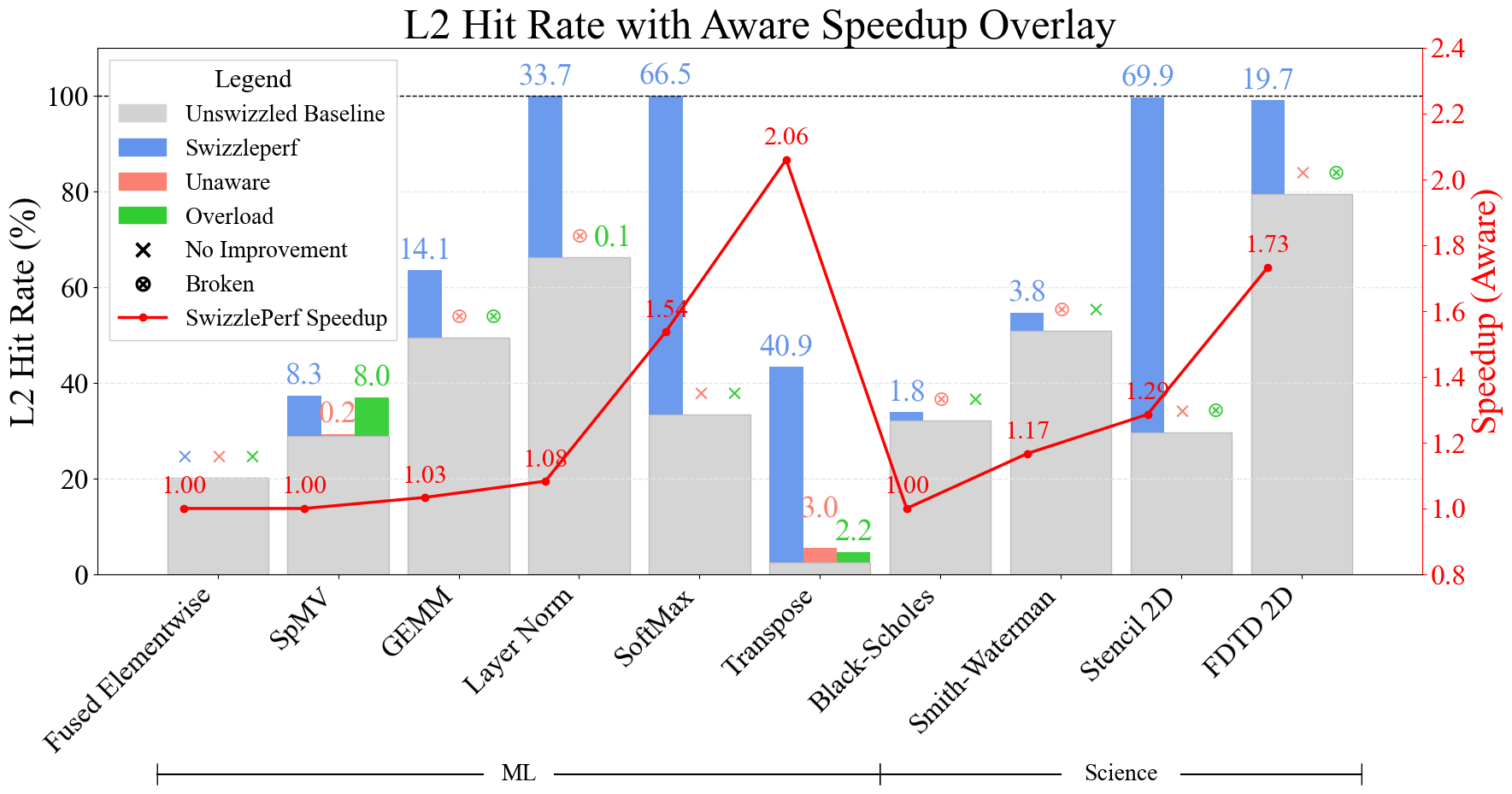}
\caption{\textit{L2 hit rate improvements and speedups from swizzling patterns on 10 kernels.} The \textcolor{gray}{gray bars} denote the original L2 hit rate of unswizzled code, \textcolor{blue}{blue bars} show the L2 hit rate improvement from the SwizzlePerf-generated swizzling patterns, and \textcolor{orange}{orange} and \textcolor{ForestGreen}{green bars} show baselines. The \textcolor{red}{red line} denotes the corresponding SwizzlePerf end-to-end kernel speedup. SwizzlePerf achieves speedups on 9 out of 10 kernels of up to $2.06\times$. These speedups are driven by higher cache efficiency, where SwizzlePerf improves L2 hit rate by an average of 23.9\% and up to 70\%. 4 of these SwizzlePerf implementations approach the maximum cache efficiency of 100\% L2 hit rate.}
\label{results}

\end{figure}

We compare the hardware-aware \textit{SwizzlePerf} against two baselines: \textit{Hardware-unaware}, the base IntelliPerf loop without hardware or scheduling context, and \textit{Hardware-overload}, which passes in an unfiltered 10k+ token public GPU architecture documentation dump. We evaluate 10 GPU kernels that are relevant to real-world workloads. We collect 6 ML kernels (GEMM, fused elementwise, layer normalization, softmax, naive sparse matrix vector multiplication (SpMV), transpose) and 4 Science workloads (Black-Scholes, finite-difference time-domain (FDTD) 2D, Smith-Waterman, Stencil 2D). We benchmark on medium problem sizes (\textasciitilde5ms) and validate correctness against reference implementations (PyTorch for ML kernels, CPU implementation for Science kernels). In Figure~\ref{results}, $\times$ denotes cases where swizzling had no effect on L2 hit rate, while \nonbijective marks broken remappings.

Across the 10 kernels, SwizzlePerf achieves an average speedup of $1.29\times$ and up to a $2.06\times$ on the transpose kernel. This large gain on transpose comes from striding the M×N tile grid across XCDs so that both the original reads and the transposed writes stay within the same XCDs’s L2 cache, eliminating cross-XCD thrashing. The softmax kernel has a $1.54\times$ speedup by grouping all row chunks into the same XCD across its two-phase reduction, which keeps row values resident in L2 and reduces conflict misses. These results show that SwizzlePerf-generated patterns consistently improve end-to-end runtimes and overall system efficiency on disaggregated GPUs.

While our swizzling results show strong end-to-end speedups, we can take advantage of our direct optimization for cache locality. It is more representative to measure success through L2 hit rate rather  runtime alone. Runtime can be noisy due to kernel launch overheads, compute saturation, and overlapping bottlenecks, but L2 hit rate directly reflects whether cache-aware remapping is successful. SwizzlePerf finds patterns that improve L2 hit rates by up to 70\%, with an average of 23.9\%. The swizzled implementation of 4 kernels reach nearly 100\% L2 hit rate, showing how SwizzlePerf can help approach the hardware maximum for cache efficiency.

For compute-bound kernels like GEMM, the 14\% L2 hit rate improvement translates to a modest $1.03\times$ speedup. Conversely, memory-sensitive kernels like stencil and transpose show that when L2 locality is the bottleneck, hit rate improvements enable significant speedups. Strong gains across the board showcase the importance of \textit{intentionally} providing relevant hardware context regardless of the kernel's specific bottleneck. By contrast, the hardware-unaware and hardware-overload swizzling patterns lead to minimal L2 hit rate improvements and never give a speedup.

Hardware-awareness and specific guiding metrics enable SwizzlePerf to reliably uncover optimizations that both (a) validate the generalizable effectiveness of swizzling (Appendices \ref{appendix_prog} and \ref{appendix_problemsize}) and (b) translate into significant end-to-end speedups.

\vspace{-8pt}

\section{Discussion and Future Work}
\vspace{-4pt}

SwizzlePerf addresses the missing hardware-awareness in today’s autonomous performance engineering frameworks. By demonstrating that an LLM equipped with hardware-aware context can replicate expert reasoning in minutes on a wide range of GPU kernels, we underscore how closing the hardware–software feedback loop unlocks substantial efficiency gains. Looking ahead, we believe the next breakthrough will come from expanding the modalities through which an LLM perceives hardware to improve kernel performance and power \cite{mlperfpower}. Non-text modalities like visualizations of swizzling patterns are challenging because they can be algorithm-specific. We believe that attention kernels are a great place to start because they are functionally similar to many different implementations \cite{flashattention, leanattention, gqa, mha, flexattention}, each with its own optimal swizzling pattern.

Our future work is guided by the question - \textbf{What are the right modalities of hardware-awareness that enable LLMs to think like a human performance engineer on real-world workloads?}

\appendix

\section{Technical Appendices and Supplementary Material}

\lstset{%
  basicstyle=\ttfamily\footnotesize,
  breaklines=true,
  columns=fullflexible,
  frame=single,
  aboveskip=0pt,   
  belowskip=0pt    
}

\newtcolorbox{fullwidthbox}[1]{%
  title=\textbf{\textit{#1}},  
  colframe=black!35,           
  colback=white,
  boxrule=0.4pt,
  arc=4pt,                     
  left=8pt,right=8pt,top=6pt,bottom=6pt,
  width=\linewidth             
}

\subsection{LLM Input Prompt and Output Structure}\label{appendix_prompts}

\begin{figure}[H]
  \centering
  \begin{fullwidthbox}{Input prompt}
    \RaggedRight
    \setstretch{1.5}   

    The original code is \{...\} with bottleneck \{...\}

    The memory analysis is \{...\}

    History of previous optimization attempts (do not repeat an implementation):
    \{\newline
    \quad iteration: 1, applied diff: \{...\}, bottleneck report: \{...\},\\
    \quad iteration: N-1, applied diff: \{...\}, bottleneck report: \{...\}
    \newline\}

    On the AMD MI300X, there are \{A\} XCDs, each has \{B\} CUs and a \{C\} MB L2 cache

    Blocks are scheduled \{Round-robin to XCDs\}

    Your swizzling goal is to \{...\}, pay special attention to \{...\}, code should be structured as \{...\}
  \end{fullwidthbox}
  \caption{Structure of hardware-awareness input prompt to LLM.}
  \label{fig:input-prompt}
\end{figure}

\vspace{-4pt}

Figure \ref{fig:input-prompt} shows the structure of the code optimization input prompt. Note that it takes in the original unoptimized code with its bottleneck report, the memory analysis from the LLM memory analysis call, the history of all prior iterations and bottlenecks, and a description of the hardware architecture with profiled details. 

\begin{figure}[H]
  \centering
  \begin{fullwidthbox}{Output Signature}
    \RaggedRight
    \setstretch{1.5}

    {Chain-of-Thought Reasoning}

    JSON dict of why old implementations were suboptimal:
    \{\newline
    \quad iteration 1: This implementation only marginally increased L2 hit rate because \{...\},\\
    \quad iteration N-1: This iteration didn’t change L2 hit rate because \{...\}
    \newline\}

    My new swizzling approach will remap blocks by \{...\}

    My new swizzling approach will be better than any prior implementation because \{...\}

    {Final code}
  \end{fullwidthbox}
  \caption{Structure of DSPy output signature.}
  \label{fig:output-signature}
\end{figure}

\vspace{-6pt}

Figure \ref{fig:output-signature} shows the structure of the code optimization output signature. The output signature is a parameter into DSPy that structures how we want the LLM to output information. We first output an unstructured chain-of-thought reasoning trace, and then explicitly reason about the shortcomings about each prior iteration. Lastly, we rationale the new implementation and output the code for it.

\subsection{Case Study: General Matrix Multiplication (GEMM) Swizzling}\label{appendix_gemm}
\begin{figure}[htbp]
  \centering
  \begin{subfigure}{0.23\textwidth} 
    \centering
    \includegraphics[width=\linewidth]{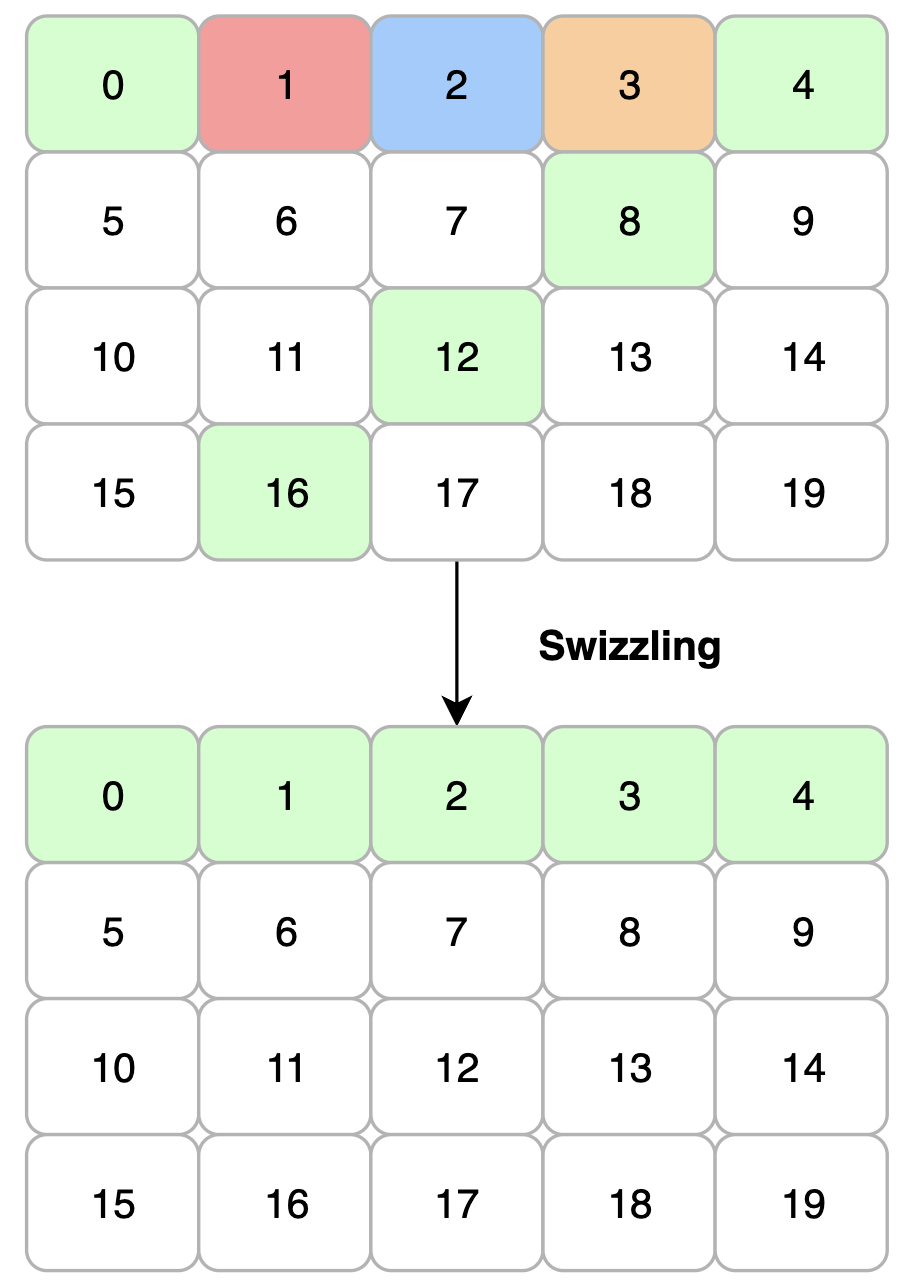}
    \caption{GEMM swizzling.}
  \end{subfigure}\hfill
  \begin{subfigure}{0.74\textwidth} 
    \centering
\begin{lstlisting}[language=Python]
pid = tl.program_id(0)

num_xcds   = 8
num_blocks = NUM_SMS

# Calculate blocks per XCD with ceiling division
b_per_xcd = (num_blocks + num_xcds - 1) // num_xcds

# Swizzle so all contiguous blocks are on same XCD
# before moving to the next XCD
pid = (pid % num_xcds) * b_per_xcd + (pid // num_xcds)
\end{lstlisting}
    \caption{SwizzlePerf-generated swizzling Triton code. Note the additional edge case catch with ceiling division.}
  \end{subfigure}

  \caption{GEMM swizzling strategy and equivalent SwizzlePerf-generated swizzling pattern.}
  \label{gemmswizzle_appendix}
\end{figure}

In tiled GEMM implementations, tiles closer together in the output matrix C (M$\times$N) share more data from input matrices A (M$\times$K) and B (K$\times$N). Tiles in the same row in C share the rows from A, and tiles in the same column of C share columns from B. Unfortunately, because of the default round-robin block scheduling in the M=4, N=5, \#XCD=4 example in Figure \ref{gemmswizzle_appendix}, adjacent tiles are placed on different XCDs and we therefore observe a hit rate on the L2 cache shared by each XCD below the theoretical maximum. While we cannot explicitly reschedule blocks to new CUs, we can make each CU process different input data and write to a different output tile by recomputing the PIDs of each block. This essentially trades work between CUs, has minimal overhead to compute, and can improve L2 hit rate in many kernels.

SwizzlePerf automatically generates swizzling patterns for any kernel algorithm, hardware architecture, or default block scheduling scheme. The GEMM swizzling example in Figure \ref{gemmswizzle_appendix} took expert performance engineers 2 weeks to design and validate. SwizzlePerf achieved a functionally identical solution with additional edge case catching in < 5 minutes. The ceiling division edge case catch when calculating blocks per XCD is not present in the expert-generated swizzling pattern, and implies that SwizzlePerf is applying new techniques and not just retrieving existing solutions.

\vspace{-15pt}

\subsection{Case Study: Other Swizzling Pattern Examples and Interpretations}\label{appendix_examples}

\begin{figure}[H]
  \centering
  \begin{minipage}{\textwidth}
    \begin{lstlisting}[language=Python,linewidth=\textwidth]
pid_m = tl.program_id(0)
num_xcds = 8
num_blocks = M

blocks_per_xcd = (num_blocks + num_xcds - 1) // num_xcds
xcd_id = (pid_m // num_xcds) % num_xcds

local_block_id = pid_m % num_xcds
offset = (local_block_id // num_xcds) * num_xcds + xcd_id

pid = offset * blocks_per_xcd + local_block_id
    \end{lstlisting}
  \end{minipage}
  \caption{SwizzlePerf-generated swizzling pattern for layer normalization kernel.}
  \label{fig:swizzle_layernorm}
\end{figure}
\vspace{-6pt}

In the layernorm swizzling shown in Figure \ref{fig:swizzle_layernorm}, we remap pid\_m so that all BLOCK\_SIZE\_N column-chunks for the same input row are routed to the same XCD. The kernel keeps that row’s partial sums (mean and variance) and the loaded weight/bias vectors resident in that XCD’s L2 across multiple loads, massively boosting intra‐row data reuse. At the same time, spreading different rows evenly over all eight XCDs prevents any single L2 cache from thrashing, which improves hit rate by the observed 34\%.

\begin{figure}[H]
  \centering
  \begin{minipage}{\textwidth}
    \begin{lstlisting}[language=Python,linewidth=\textwidth]
num_xcds = 8
num_blocks = tl.num_programs(0)

pid = tl.program_id(0)

blocks_per_xcd = (num_blocks + num_xcds - 1) // num_xcds
xcd_id = (pid // num_xcds) % num_xcds
block_in_xcd = pid % num_xcds
new_pid = xcd_id * blocks_per_xcd + block_in_xcd
pid = new_pid

    \end{lstlisting}
  \end{minipage}
  \caption{SwizzlePerf-generated swizzling pattern for softmax kernel.}
  \label{fig:swizzle_softmax}
\end{figure}
\vspace{-6pt}

In the softmax swizzling shown in Figure \ref{fig:swizzle_softmax}, because softmax for each row does a two‐phase reduction (first finding the maximum, then exponentiating and summing), grouping all the BLOCK\_SIZE chunks of a given row into the same XCD keeps those row values resident in that XCDs’s L2 cache across both passes, rather than bouncing them through DRAM. At the same time, by evenly interleaving different rows across all eight XCDs, no single XCD becomes a hotspot, so conflict misses and evictions drop—together driving a 66\% lift in L2 hit rate.
\vspace{-6pt}

\begin{figure}[H]
  \centering
  \begin{minipage}{\textwidth}
    \begin{lstlisting}[language=Python,linewidth=\textwidth]
pid_x = tl.program_id(0)
pid_y = tl.program_id(1)

num_XCD = 8
num_blocks_x = (Nx + BLOCK_SIZE_X - 1) // BLOCK_SIZE_X
num_blocks_y = (Ny + BLOCK_SIZE_Y - 1) // BLOCK_SIZE_Y
num_blocks = num_blocks_x * num_blocks_y

# Calculate new block index using a different swizzling pattern
block_id = pid_y * num_blocks_x + pid_x
xcd_id = (block_id // num_blocks_x) % num_XCD
new_block_id = (block_id % num_blocks_x) * num_XCD + xcd_id

# Map new block id back to 2D grid
pid_x = new_block_id // num_blocks_x
pid_y = new_block_id % num_blocks_x
    \end{lstlisting}
  \end{minipage}
  \caption{SwizzlePerf-generated swizzling pattern for finite-difference time-domain kernel.}
  \label{fig:swizzle_fdtd}
\end{figure}
\vspace{-8pt}

In the finite difference time domain (FDTD) swizzling shown in Figure \ref{fig:swizzle_fdtd}, it maps each 2D FDTD block so that blocks lying on the same vertical stride (i.e., sharing the same x-range but different y-ranges) end up on the same XCD. At the same time, by distributing those strides round-robin across all eight XCDs (via the xcd\_id = pid\_y \% num\_XCD step), no gets overloaded with too many neighbor-dependent blocks, so you avoid cache thrashing. Together, this alignment of FDTD’s neighboring accesses and balanced loads across XCDs improves L2 hit rate by 20\%.

\begin{figure}[H]
  \centering
  \begin{minipage}{\textwidth}
    \begin{lstlisting}[language=Python,linewidth=\textwidth]
pid_m = tl.program_id(0)
pid_n = tl.program_id(1)

num_blocks_m = (M + BLOCK_SIZE_M - 1) // BLOCK_SIZE_M
num_blocks_n = (N + BLOCK_SIZE_N - 1) // BLOCK_SIZE_N
total_blocks = num_blocks_m * num_blocks_n

block_id = pid_m * num_blocks_n + pid_n
xcd_id = (block_id // num_blocks_m) % 8
round_id = block_id // (8 * num_blocks_m)
new_block_id = round_id * 8 + xcd_id

pid_m = new_block_id // num_blocks_n
pid_n = new_block_id % num_blocks_n
    \end{lstlisting}
  \end{minipage}
  \caption{SwizzlePerf-generated swizzling pattern for stencil 2D kernel.}
  \label{fig:swizzle_stencil}
\end{figure}

In the stencil 2D swizzling shown in Figure \ref{fig:swizzle_stencil}, it remaps the blocks so that blocks which are neighbors in the M or N direction (and thus share data for the dimensions in stencil) end up on the same XCD’s L2 cache. By laying out the grid in these diagonal-stride groups, each XCD keeps its center-and-neighbor loads resident across multiple block invocations, while the round-robin scheduling prevents any one XCD from thrashing-together, and improves L2 hit rate by 70\%.

\begin{figure}[H]
  \centering
  \begin{minipage}{\textwidth}
    \begin{lstlisting}[language=Python,linewidth=\textwidth]
num_xcds = 8
pid_m = tl.program_id(0)
pid_n = tl.program_id(1)

pid = pid_m * (N // BLOCK_SIZE_N) + pid_n
xcd_idx = (pid // num_xcds) % num_xcds
pid = xcd_idx * (N // BLOCK_SIZE_N) * (M // BLOCK_SIZE_M) // num_xcds + pid // num_xcds

pid_m = pid // (N // BLOCK_SIZE_N)
pid_n = pid % (N // BLOCK_SIZE_N)
    \end{lstlisting}
  \end{minipage}
  \caption{SwizzlePerf-generated swizzling pattern for transpose kernel.}
  \label{fig:swizzle_transpose}
\end{figure}
\vspace{-6pt}

In the transpose swizzling shown in Figure \ref{fig:swizzle_transpose}, it strides the linearized M×N tile grid across the eight XCDs. This takes each block’s row‐major index modulo num\_xcds to pick a XCD and then re-assembles the remaining blocks contiguously. Each XCD ends up owning whole “bands” of input rows and their corresponding output columns. For transpose, that means when you load a BLOCK\_SIZE\_M×BLOCK\_SIZE\_N tile and then write its BLOCK\_SIZE\_N×BLOCK\_SIZE\_M transpose, both the original reads and the transposed writes hit the same XCD’s L2 rather than thrashing across XCDs. At the same time, because the blocks are evenly distributed, no single L2 becomes a hotspot and leads to a 34\% improvement in hit rate.

\subsection{Case Study: Hardware-Overload Induces Bitwise Swizzling}\label{appendix_bitwise}

\begin{figure}[H]
  \centering
  \begin{minipage}{\textwidth}
    \begin{lstlisting}[language=Python,linewidth=\textwidth]
pid_m = tl.program_id(0)
pid_n = tl.program_id(1)

num_xcds = 8
num_blocks_m = (M + BLOCK_SIZE_M - 1) // BLOCK_SIZE_M
num_blocks_n = (N + BLOCK_SIZE_N - 1) // BLOCK_SIZE_N
total_blocks = num_blocks_m * num_blocks_n

original_block_index = pid_m * num_blocks_n + pid_n

xcd_id = original_block_index % num_xcds
round_id = original_block_index // num_xcds
new_block_index = round_id + xcd_id * (total_blocks // num_xcds)

pid_m = new_block_index // num_blocks_n
pid_n = new_block_index % num_blocks_n
    \end{lstlisting}
  \end{minipage}
  \caption{Hardware-aware SwizzlePerf swizzling code for transpose kernel.}
  \label{fig:swizzle_hw_aware}
\end{figure}

\begin{figure}[H]
  \centering
  \begin{minipage}{\textwidth}
    \begin{lstlisting}[language=Python,linewidth=\textwidth]
pid = tl.program_id(0)

# Swizzle the program IDs to improve cache locality
pid_m = pid // (N // BLOCK_SIZE_N)
pid_n = pid % (N // BLOCK_SIZE_N)
    \end{lstlisting}
  \end{minipage}
  \caption{Hardware-unaware swizzling code for transpose kernel.}
  \label{fig:swizzle_unaware}
\end{figure}

\begin{figure}[H]
  \centering
  \begin{minipage}{\textwidth}
    \begin{lstlisting}[language=Python,linewidth=\textwidth]
pid = tl.program_id(0)

# Swizzle the pid to improve cache locality
pid = ((pid >> 1) & 0x55555555) | ((pid & 0x55555555) << 1)
    \end{lstlisting}
  \end{minipage}
  \caption{Hardware-overload swizzling code for transpose kernel.}
  \label{fig:swizzle_overload}
\end{figure}

\vspace{-15pt}

In experiments on specific kernels and problem sizes, the hardware-overload optimization loop achieves strong speedups by applying bitwise SHIFT and ADD swizzling (Figure~\ref{fig:swizzle_overload}). Hardware-unaware applies a simple size-based pattern (Figure~\ref{fig:swizzle_unaware}), while SwizzlePerf generates more complex, MI300x-specific mappings (Figure~\ref{fig:swizzle_hw_aware}). The bitwise approach can achieve up to 70\% higher L2 hit rate than SwizzlePerf, but it is overfit to particular sizes and often fails correctness. For example, on the transpose kernel with M=32768 and N=32768, a low-bit swap aligned perfectly to co-locate tiles on the same XCD, but similar patterns failed on other kernels.

These results show that hardware-overload swizzling does not generalize: if the tile count is not aligned with the number of XCDs, if BLOCK\_SIZE\_M/N changes, or if the GPU has a different XCD count, the hard-coded swap breaks reuse. Many overload patterns also failed bijectivity, sending blocks to non-existent partners when grid shapes changed. While hardware-overload offers an interesting idea that could be integrated into SwizzlePerf, its lack of robustness across problem sizes limits usability. By contrast, SwizzlePerf generalizes correctly across kernels and sizes, even if it does not always achieve the very best result on every configuration.

\vspace{-5pt}

\subsection{Optimization Loop Progression}\label{appendix_prog}

\begin{figure}[htbp]
  \centering

  \subcaptionbox{GEMM kernel progression.\label{fig:gemm_prog}}[0.49\linewidth]{%
    \includegraphics[width=\linewidth]{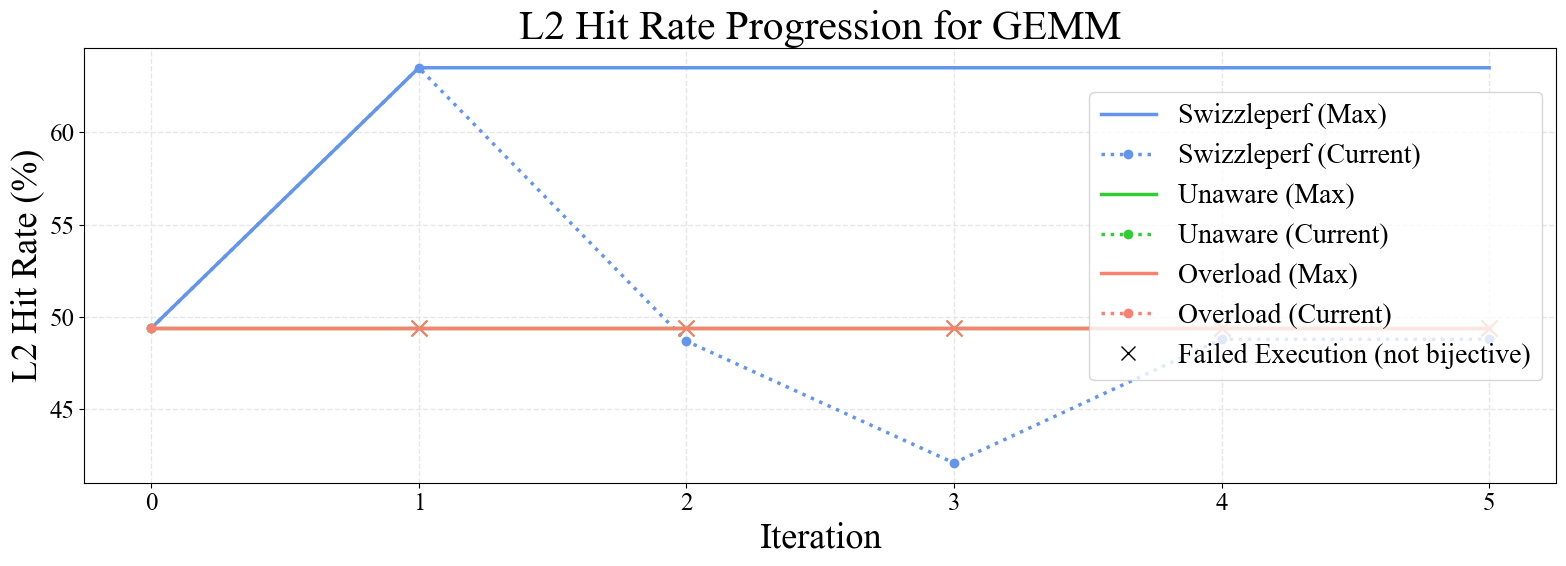}}
  \hfill
  \subcaptionbox{Stencil 2D kernel progression.\label{fig:stencil_prog}}[0.49\linewidth]{%
    \includegraphics[width=\linewidth]{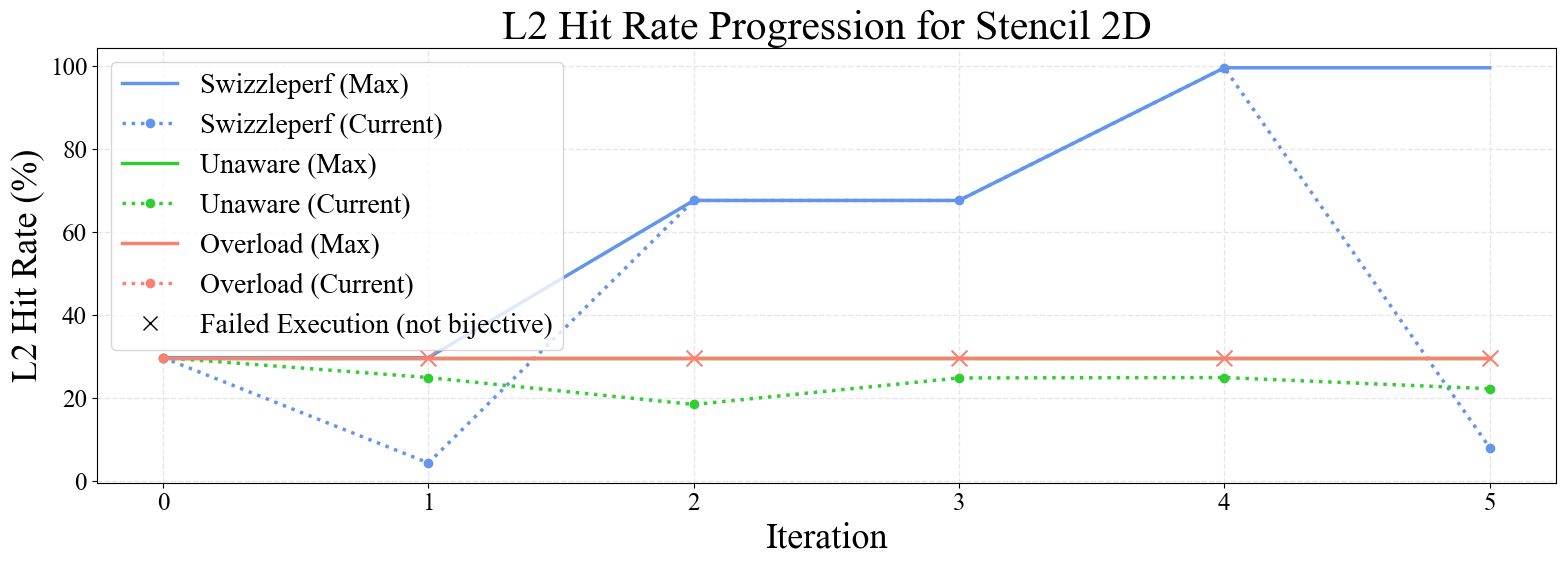}}
  \hfill
  \subcaptionbox{Naive SpMV kernel progression.\label{fig:spmv_prog}}[0.49\linewidth]{%
    \includegraphics[width=\linewidth]{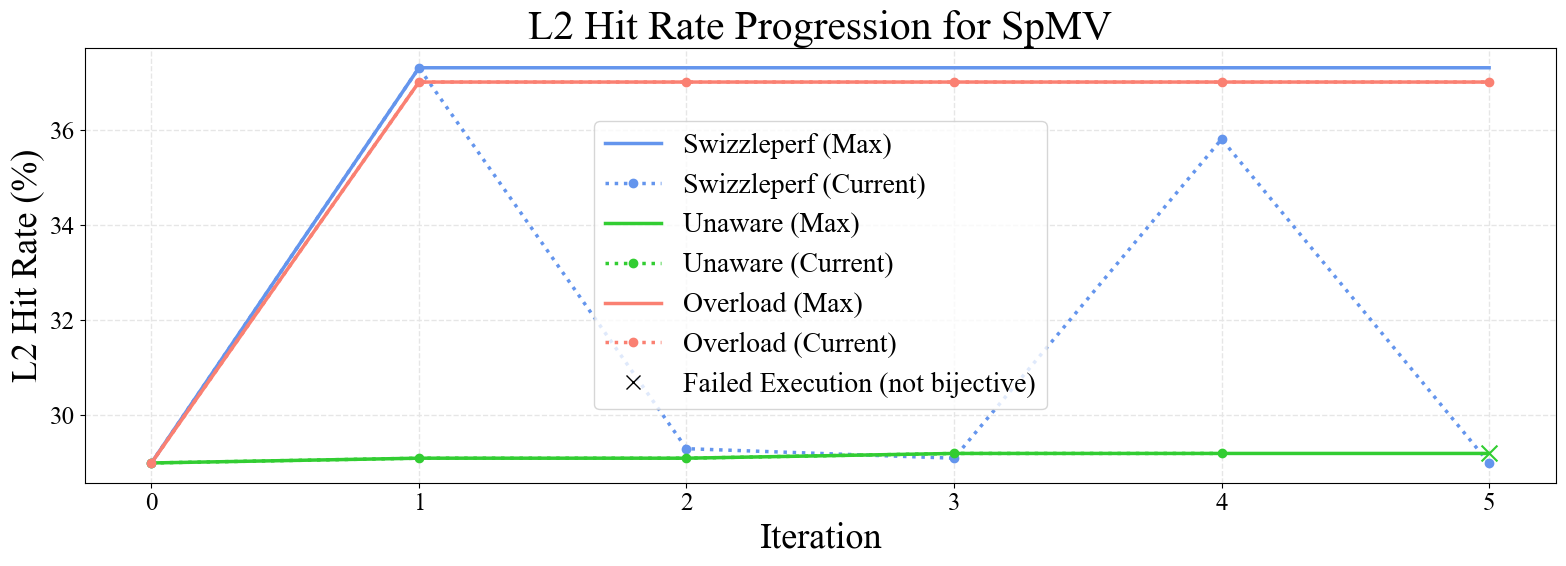}}

  \subcaptionbox{Softmax kernel progression.\label{fig:softmax_prog}}[0.49\linewidth]{%
    \includegraphics[width=\linewidth]{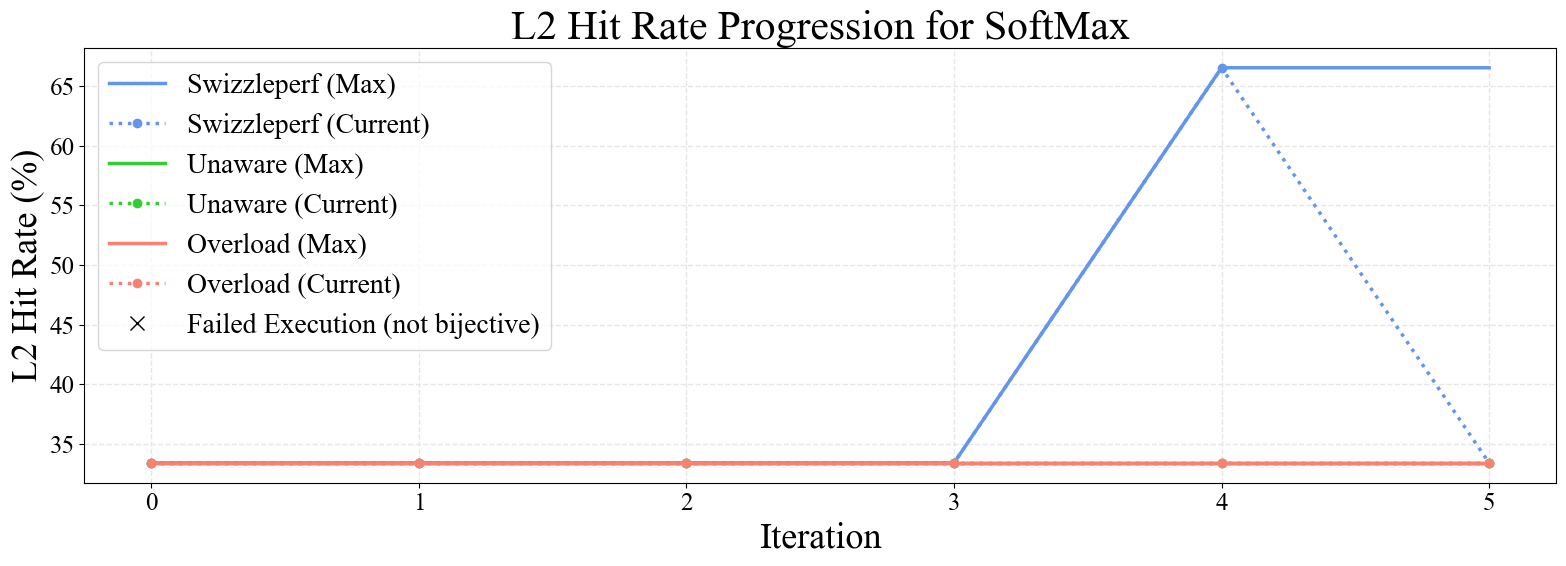}}
  \hfill
  \subcaptionbox{LayerNorm kernel progression.\label{fig:layernorm_prog}}[0.49\linewidth]{%
    \includegraphics[width=\linewidth]{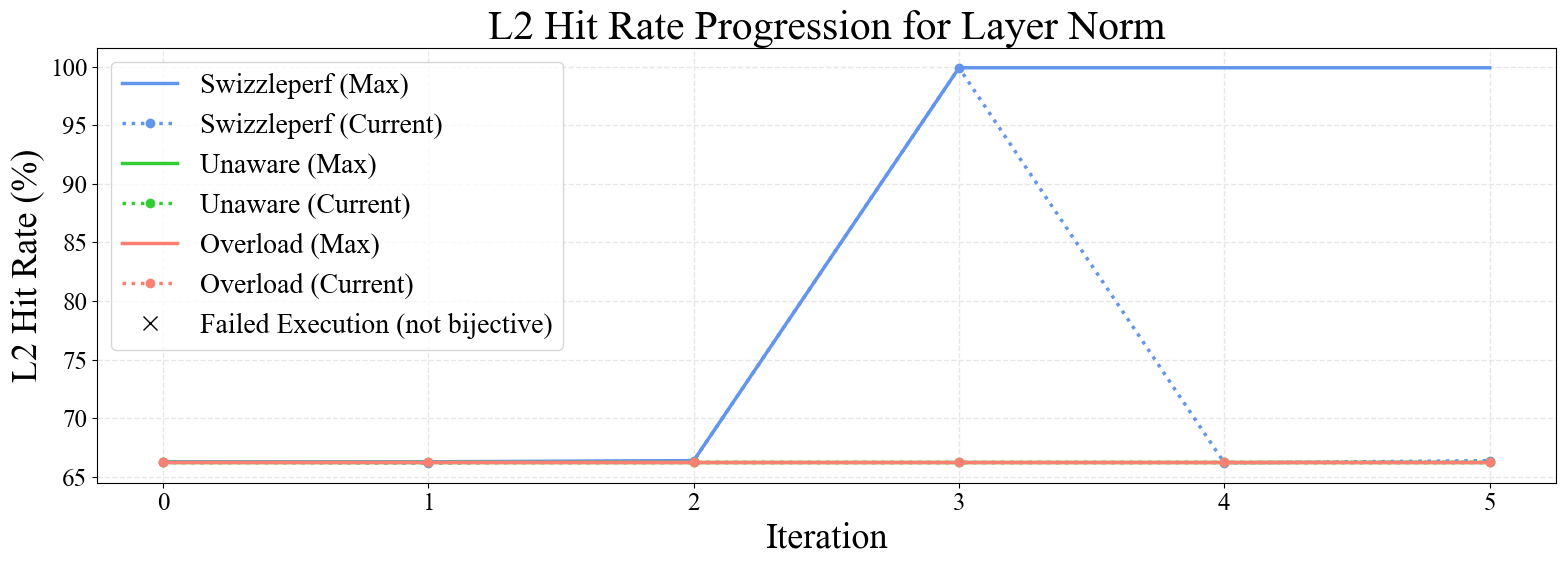}}
  \hfill
  \makebox[0.32\linewidth]{} 

  \caption{Progression plots for GEMM, Stencil 2D, SpMV, Softmax, and LayerNorm kernels. SwizzlePerf consistently finds more performant swizzling patterns and explores more diverse solutions than baselines. The dotted line shows the L2 hit rate of the current iteration’s implementation, while the solid line tracks the best-so-far implementation.}
  \label{fig:progression_all}
\end{figure}

We evaluate the swizzling patterns generated in subsequent iterations of SwizzlePerf against the two baselines. In Figure~\ref{fig:progression_all}, the dotted line is the L2 hit rate of the implementation from the numbered iteration, and the solid line is the best prior implementation that would be returned. Each iteration receives context on the weaknesses of earlier swizzling patterns, so improvement in hit rate represents successful reflection on past approaches. 

For GEMM in Figure~\ref{fig:gemm_prog}, SwizzlePerf finds a correct swizzling pattern within one iteration, while both baselines fail across all five iterations. For Stencil 2D in Figure~\ref{fig:stencil_prog}, SwizzlePerf shows steady improvement to nearly 100\% L2 hit rate and explores diverse solutions instead of getting stuck in a local minimum. The hardware-unaware approach consistently generates patterns that degrade L2 hit rate, and the hardware-overload approach always fails by generating incorrect swizzling patterns with a bitwise AND/SHIFT strategy (Appendix~\ref{appendix_bitwise}).  

For Naive SpMV in Figure~\ref{fig:spmv_prog}, hardware-overload achieves slightly worse performance than SwizzlePerf and fixates on its initial method instead of exploring alternatives. For Softmax in Figure~\ref{fig:softmax_prog} and LayerNorm in Figure~\ref{fig:layernorm_prog}, SwizzlePerf converges on successful patterns after multiple iterations, while the baselines fail to adapt.  

\subsection{Ablation Study: Evaluating on Different Problem Sizes}\label{appendix_problemsize}

We evaluate the SwizzlePerf generated swizzling patterns across different problem sizes of the target kernel. It is important to ensure that the implementations are generalizable to many inputs and not just overfit to a specific problem size.

\begin{figure}[H] 
  \centering
  \begin{subfigure}[b]{0.48\linewidth}
    \centering
    \includegraphics[width=\linewidth]{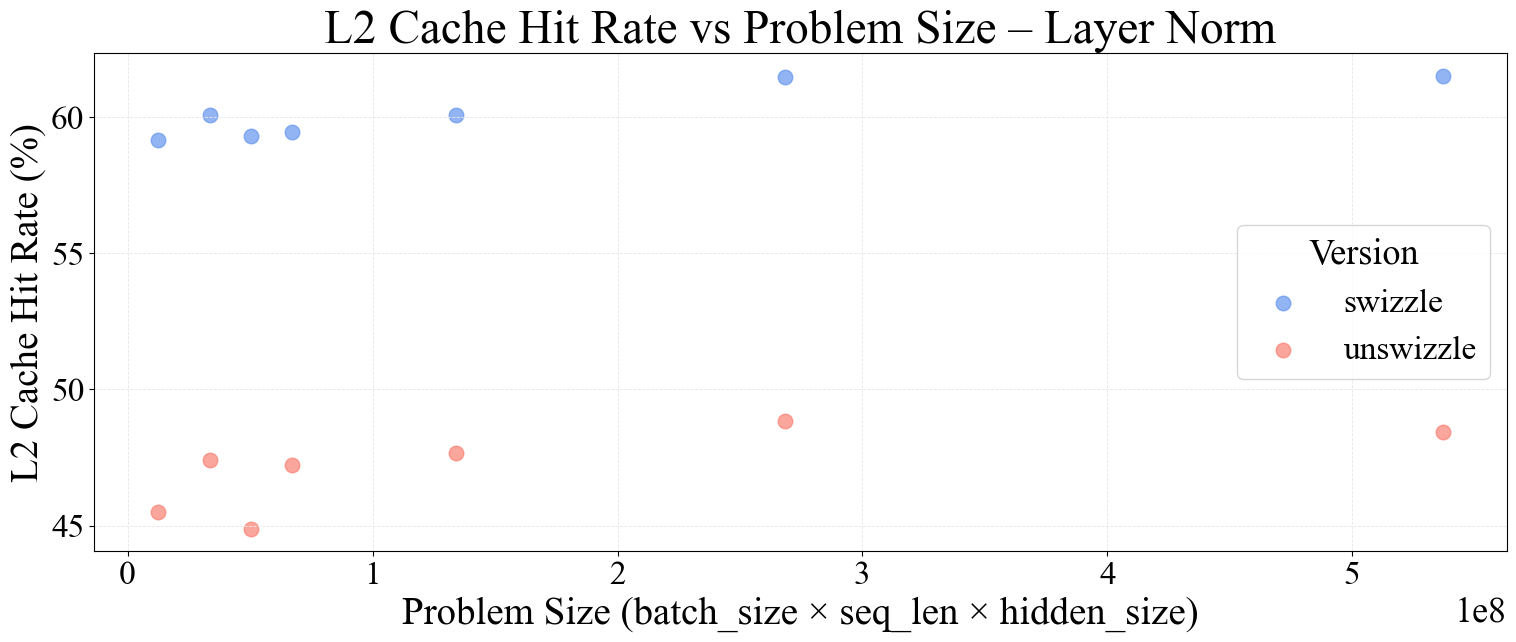}
    \caption{LayerNorm}
    \label{fig:abl_layernorm}
  \end{subfigure}
  \begin{subfigure}[b]{0.48\linewidth}
    \centering
    \includegraphics[width=\linewidth]{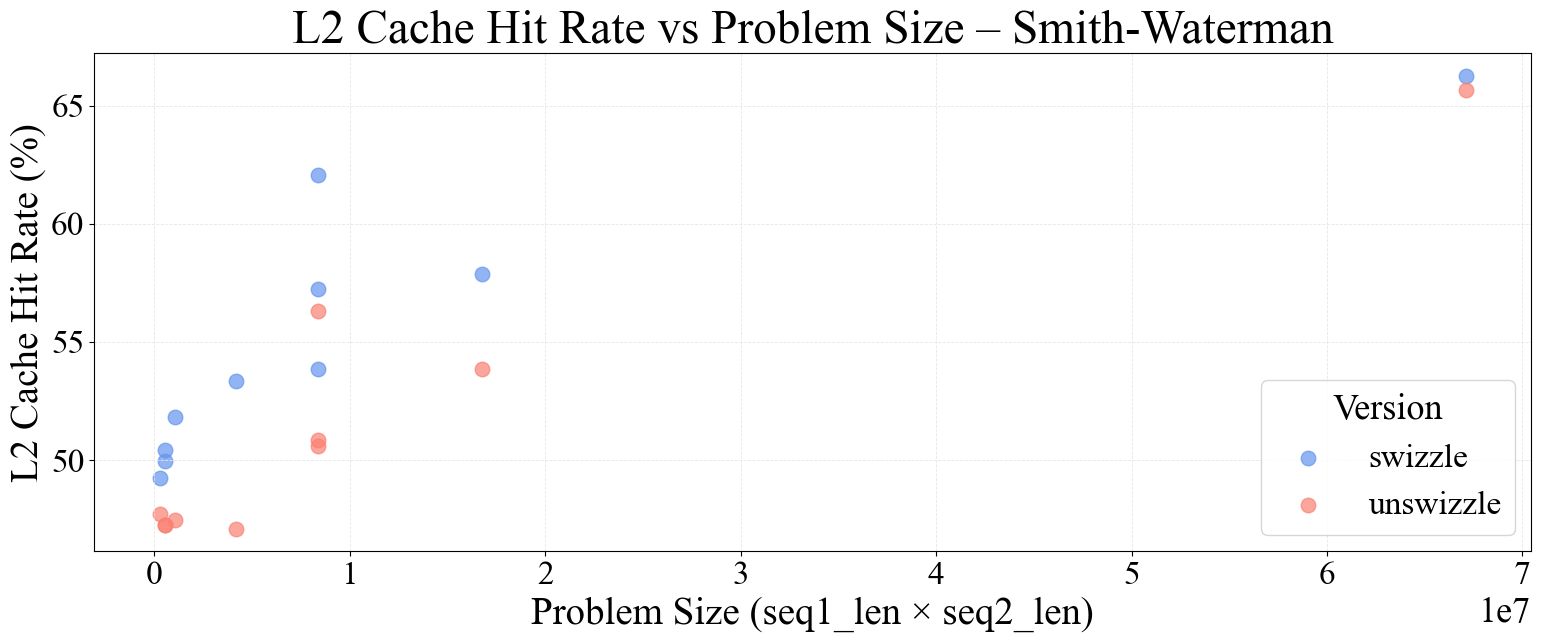}
    \caption{Smith–Waterman}
    \label{fig:abl_sw}
  \end{subfigure}
  \begin{subfigure}[b]{0.48\linewidth}
    \centering
    \includegraphics[width=\linewidth]{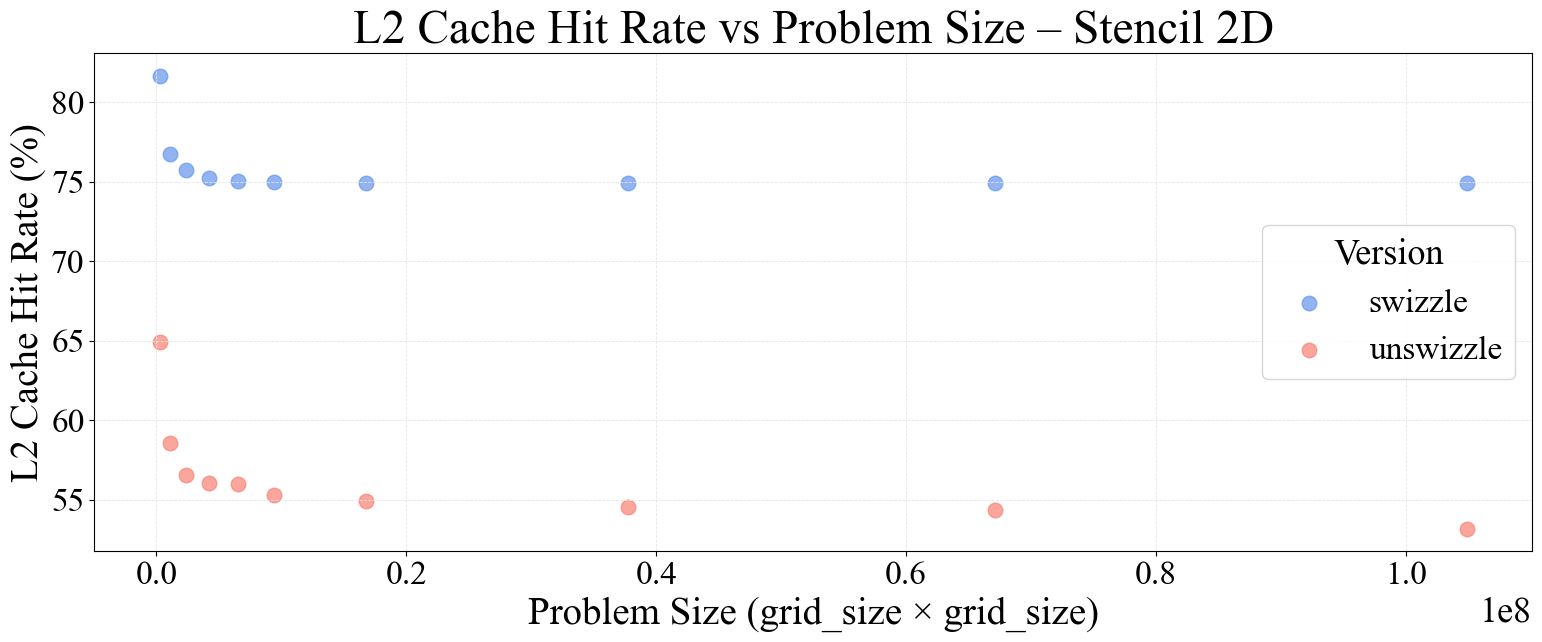}
    \caption{Stencil 2D}
    \label{fig:abl_stencil}
  \end{subfigure}
  \vspace{-0.7em}
  \caption{\small L2 cache hit rate vs.\ problem size for the three kernels.}
  \label{fig:abl_problem_size}
\end{figure}

The layernorm kernel shown in Figure ~\ref{fig:abl_layernorm} is evaluated on problem sizes $64\times1024\times1024$ to $512\times8192\times1024$ elements.
Swizzling lifts the L2 hit rate from 46\% to 60\%. This delta remains essentially flat as the
tensor grows. The consistent gap indicates that SwizzlePerf was able to find a pattern that was generalizable and consistently outperforms the unswizzled for.

The Smith-Waterman kernel shown in Figure \ref{fig:abl_sw} is evaluated across five doublings in sequence length (from \(512^{2}\) to \(8192^{2}\) DP cells), the swizzled kernel raises the L2 hit rate from 50\% to 65\%, while the unswizzled baseline tops out near \(56\%\).  The gap therefore widens from about 3\% on the smallest input to 10\% on the largest. This equates to about a 15-20\% relative gain that becomes more pronounced as the grid grows.

Stencil 2D kernel shown in Figure \ref{fig:abl_stencil} is evaluated with grid side increasing twenty-fold (\(512\times512\) to \(10{,}240\times10{,}240\)
cells). The unswizzled hit rate falls from 65\% down to 53\%, whereas the
swizzled version stays in the 74-82\% range. This yields a
20-30\% advantage at large problem sizes—over
40\% relative—showing that swizzling keeps cache locality almost size-invariant for
this Jacobi update.

\subsection{Ablation Study: Evaluating Hardware-Awareness with Different LLMs}\label{appendix_llms}

\begin{figure}[t]
    \centering
    \includegraphics[width=0.9\linewidth]{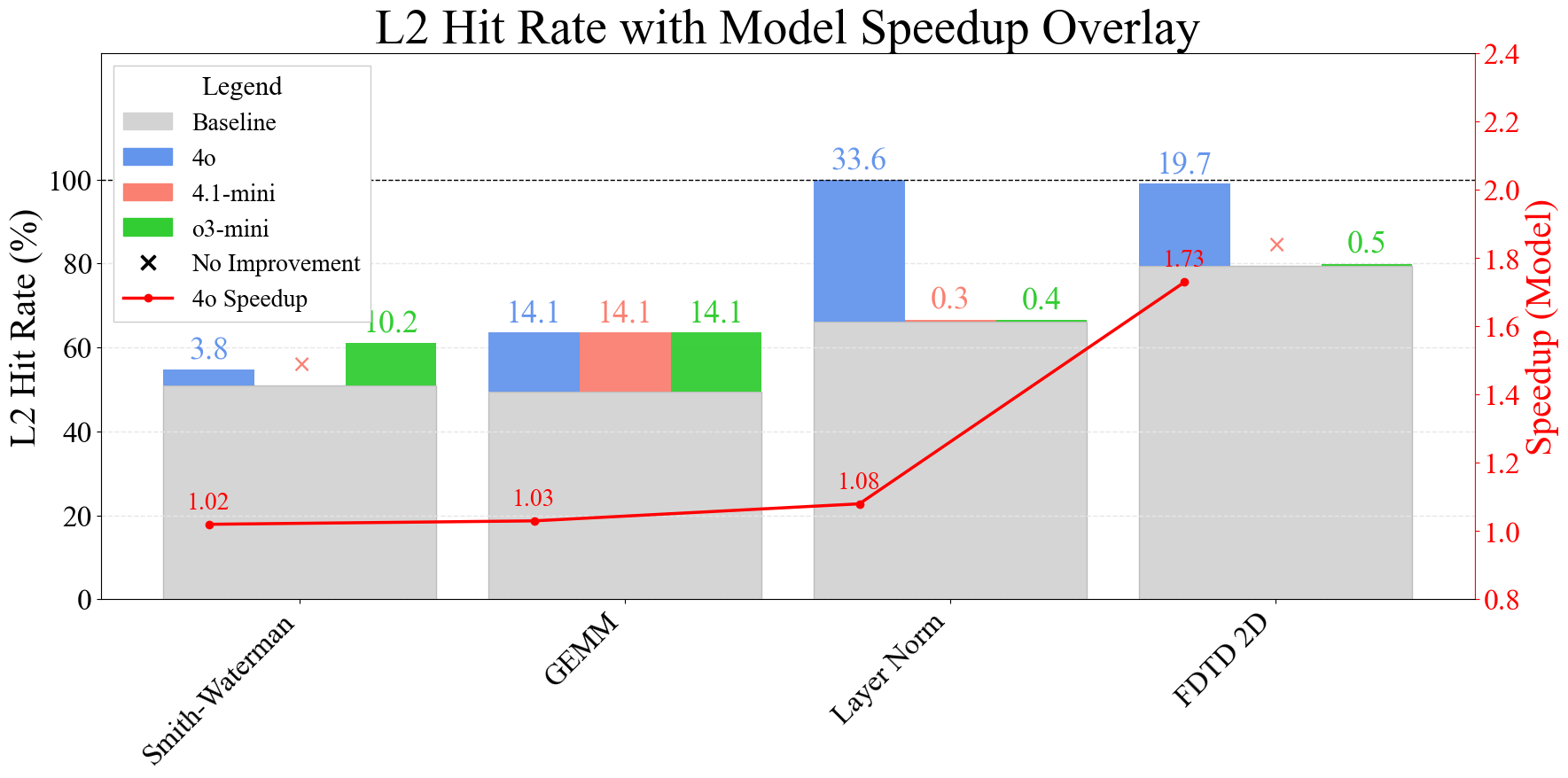}
    \caption{L2 hit rate of SwizzlePerf-generated patterns with 3 different LLMs.}\label{llm_l2}
  \end{figure}

 In Figure \ref{llm_l2}, we evaluate how different LLMs impact SwizzlePerf-generated swizzling patterns. Prior work has observed the varying capabilities on code optimization\cite{pie4perf} and computer architecture \cite{quarch} tasks. We run SwizzlePerf with OpenAI’s 4o, 4.1-mini, and o3-mini on GEMM, LayerNorm, FDTD, and Smith-Waterman.

In GEMM, all three models converge to a swizzling pattern equivalent to the expert-generated one, which is expected given GEMM’s relatively straightforward locality structure. For LayerNorm and FDTD, only 4o discovers effective solutions. 4.1-mini is likely too small to reason about complex mappings, and o3-mini (despite being a reasoning model and outperforming on most benchmarks) sometimes underperforms. In contrast, on Smith-Waterman, o3-mini generates a more complex swizzling pattern that outperforms both 4o and 4.1-mini.

These results suggest that different models excel on different kernels, and that reasoning-focused training does not always help in hardware-aware optimization where explicit context is already provided. Selecting the best model likely depends on kernel, architecture, and problem size. We plan to broaden this evaluation to open- and closed-source models across Anthropic, Meta, OpenAI, Google, etc., to better understand which approaches are most effective for hardware-awareness.

\vspace{-5pt}
\subsection{Future Consideration: Swizzling for Power Efficiency}\label{appendix_power}

While the primary focus of SwizzlePerf has been performance, we believe that same locality-aware remapping that lifts L2 hit rate will also have pronounced power-efficiency benefits. Each miss that is redirected from device memory to the on-chip L2 avoids the high energy cost of traversing the full memory hierarchy. By clustering cooperative blocks within a single XCD, our generated swizzling patterns cut off-chip traffic and stabilize residency in the disaggregated caches, trimming the average energy per instruction even in kernels whose execution time is dominated by arithmetic throughput. In short, better locality is doubly rewarded on modern chiplet GPUs, once in latency, and again in joules.

These gains persist and may even grow in ostensibly compute-bound workloads because real devices operate under dynamic voltage and frequency scaling (DVFS). More hits in the L2 lowers instantaneous power draw, pushing the kernel farther from the DVFS throttle point and permitting higher sustained clocks or headroom for co-resident kernels. Conversely, when DVFS caps frequency, swizzling can still improve energy-to-solution by ensuring a larger fraction of each watt is spent in compute instead of memory access. We can rigorously evaluate \cite{smc, hong2010integrated, weaver2012measuring} and apply these insights to make energy efficiency-driven swizzling decisions.

\section*{Acknowledgements}

This work was supported in part by Advanced Micro Devices, Inc. under the AMD AI \& HPC Cluster Program. The authors would like to thank Cole Ramos, Karthik Sangaiah, Sumanth Gudaparthi, Muhammad Osama, Michael Schulte, and Ralph Wittig for their support. AMD, the AMD Arrow logo, AMD CDNA, AMD Instinct, AMD ROCm, AMD Infinity Cache, AMD Infinity Fabric, and combinations thereof are trademarks of Advanced Micro Devices, Inc. Other product names used in this publication are for identification purposes only and may be trademarks of their respective companies.

\newpage
\bibliographystyle{plainnat}
\bibliography{swizzleperf}

\end{document}